# Structural and magnetic properties in the ruthenate Bi$_{2.67}$Pr$_{0.33}$Ru$_3$O$_{11}$

S. Zouari[*1], A. Cheikh-Rouhou[1], R. Ballou[2] and P. Strobel[3]

[1]Laboratoire de Physique des Matériaux, Faculté des Sciences de Sfax, BP 763, 3038 Sfax, Tunisie.
[2]Laboratoire Louis Néel, CNRS, B. P. 166, 38042 Grenoble, Cedex 9 France.
[3]Laboratoire de Cristallographie, CNRS, B. P. 166, 38042 Grenoble, Cedex 9 France.

**Abstract**

An unreported praseodymium substituted phase of the Bi-Ru-O system with formula Bi$_{2.67}$Pr$_{0.33}$Ru$_3$O$_{11}$ was prepared. Its crystal structure and magnetic properties were investigated. Pr substituted occurs only on one of the two Bi sites. The magnetism of the compound is essentially dominated by a Pr$^{3+}$ contribution. A non linear variation of the reciprocal magnetic susceptibility is observed that appears to account for crystal electric field interactions on the Pr$^{3+}$ ions.

## 1  Introduction

The Bi-Ru-O system includes two known phases of equal Bi/Ru ratio: the pyrochlore Bi$_2$Ru$_2$O$_7$ and another cubic compound with formula Bi$_3$Ru$_3$O$_{11}$. The latter crystallizes in the space group Pn3 [1, 2], and contains edge-sharing dimers Ru$_2$O$_{10}$ of octahedrally coordinated ruthenium, which are stacked in a 3D network. In a recent paper, Lee et al. [2] described the heat capacity, resistivity, Hall effect and magnetic susceptibility of Bi$_3$Ru$_3$O$_{11}$. The latter was ascribed to Pauli paramagnetism. In the course of a study of the Bi-Ln-Ru-O system (Ln = rare earth), we obtained for the first time a rare-earth substituted 3-3-11 phase with composition Bi$_{2.67}$Pr$_{0.33}$Ru$_3$O$_{11}$. We report here its structural and magnetic properties.

## 2  Experimental

Polycrystalline Bi-Pr-Ru-O phases were prepared using the standard ceramic processing technique by mixing Pr6O11, Bi2O3 and RuO2 up to 99.9 % purity in the desired proportions. The precursors were thoroughly mixed in an agate mortar, and the powder was pressed into pellets and repeatedly heated at 1050-1075°C with intermittent re-grinding. Annealings in air yielded an almost pure 3-3-11 phase for an initial stoichiometry Bi/Pr/Ru = 0.8/0.2/1, while annealings in dry argon flow yielded the pyrochlore phase. These results are consistent with the change in ruthenium oxidation state between these two compounds: +4.33 in the 3-3-11 phase and +4 in the pyrochlore phase.

Phase purity, homogeneity and unit cell dimensions were determined by powder X-ray diffraction at room temperature, using a Siemens D-5000 diffractometer in transmission geometry with Cu-Kα radiation. The structure was refined from room-temperature X-ray data by the Rietveld method, using the Fullprof program [3]

[*] Corresponding author: e-mail: x.y@xxx.yyy.zz, Phone: +00 999 999 999, Fax: +00 999 999 999

Magnetization measurements were performed in the temperature range 5–300 K using an axial extraction magnetometer under applied fields up to 5 T.

## 3 Results and discussion

### 3.1. Crystal chemistry

The powder diffraction pattern for $Bi_{2.67}Pr_{0.33}Ru_3O_{11}$ is shown in Figure 1. All peaks can be assigned to a 3-3-11-type cubic phase (space group Pn3) except for weak peaks revealing a minor amount of pyrochlore phase. Both phases were included in the Rietveld refinement, yielding mass ratios of 98.9(1) % 3-3-11 phase and 1.1(1) % pyrochlore phase. The fit is shown in Figure 1 and the refinement results are given in Tables 1 and 2. Selected values of the derived bond distances and angles are displayed in Table 3.

Given the low fraction of second phase, all profile parameters were constrained to be identical for phases 1 and 2. Atomic displacement parameters (ADP's) could not be significantly refined for light oxygen atoms. Because of strong correlations, ADP's and occupation factors were not refined simultaneously. The refinement of site occupations showed that the Bi1 site only (in 4b position) is substituted by Pr in a ratio 1/3Pr + 2/3Bi. We note that this site is less distorted than that of Bi2, which may explain this selective substitution. The structural data are in good agreement with the previous determination on unsubstituted $Bi_3Ru_3O_{11}$ [1], and shows a slight contraction of the cell, in agreement with the smaller ionic radius of $Pr^{3+}$ compared to $Bi^{3+}$.

The refined cell parameter of the pyrochlore phase shows that its composition is close to pure $Bi_2Ru_2O_7$ (without any praseodymium substitution).

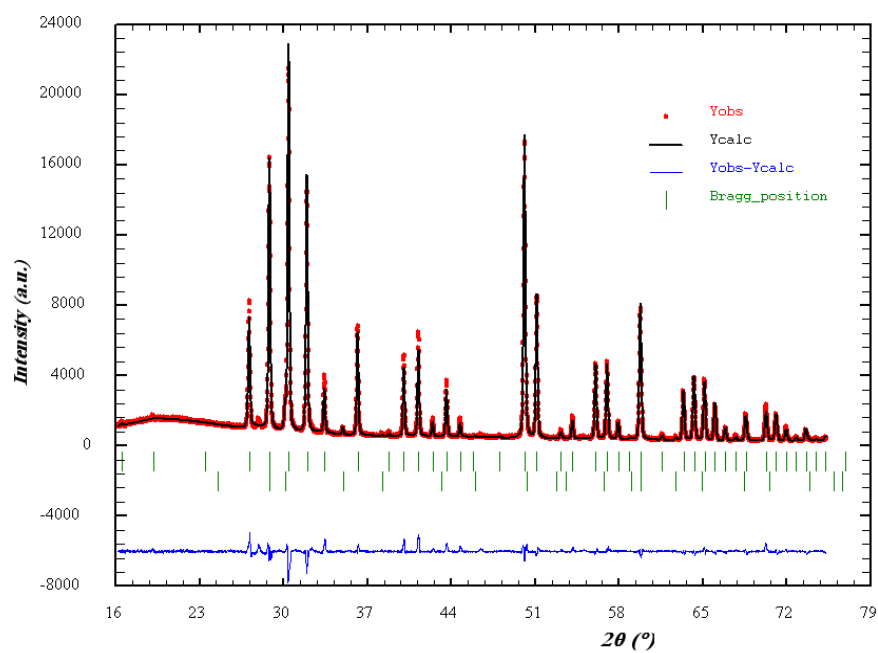

**Fig.1.** X-ray powder diffraction data for $Bi_{2.67}Pr_{0.33}Ru_3O_{11}$ at room temperature. The solid line through the data points shows the profile obtained by Rietveld analysis. The lower solid line corresponds to the difference between the fit and the data. Upper end lower vertical bars indicate the possible reflections for $Bi_{2.67}Pr_{0.33}Ru_3O_{11}$ and $Bi_2Ru_2O_7$ phases, respectively.

**Table 1.** X-ray powder refinement results for $Bi_{2.67}Pr_{0.33}Ru_3O_{11}$

| Space group | $Pn3$ |
|---|---|
| a (Å) | 9.2723(11)  ($Bi_3Ru_3O_{11}$ [1] : 9.302) |
| Line shape | Pseudo-Voigt |
| Statistical parameters | Rp 4.79, Rwp 6.44, Rexp 2.89, R-Bragg 3.98 |
| No. parameters | 54 |
| N-P+C | 3640 |
| 2θ range | 16° - 74° |

**Table 2.** Refined atomic coordinates and thermal parameters for $Bi_{2.67}Pr_{0.33}Ru_3O_{11}$

| Atom | Position | x | y | z | B(Å$^2$) | occ |
|---|---|---|---|---|---|---|
| Bi1 | 4 b | 0 | 0 | 0 | 0.586(47) | 0.112(1) |
| Pr | 4b | 0 | 0 | 0 | 0.586(47) | 0.056(1) |
| Bi2 | 8 e | 0.3835(1) | 0.3835(1) | 0.3835(1) | 1.135(35) | 0.333 |
| Ru | 12 g | 0.3903(2) | 0.75 | 0.25 | 0.634(42) | 0.5 |
| O1 | 12 f | 0.5976(16) | 0.25 | 0.25 | 1 | 0.5 |
| O2 | 8 e | 0.1500(8) | 0.1500(8) | 0.1500(8) | 1 | 0.333 |
| O3 | 24 h | 0.5885(11) | 0.2500(8) | 0.5627(13) | 1 | 1 |

**Table 3.** Select bond distances (Å) for $Bi_3Ru_3O_{11}$.[1] and $Bi_{2.67}Pr_{0.33}Ru_3O_{11}$ (this work)

|  | $Bi_3Ru_3O_{11}$ | $Bi_{2.67}Pr_{0.33}Ru_3O_{11}$ |
|---|---|---|
| $Bi_1$-$O_2$ (×2) | 2.45 | 2.4101(73) |
| $Bi_1$-$O_3$ (×6) | 2.51 | 2.5273(80) |
| $Bi_2$-$O_2$ (×3) | 2.21 | 2.2088(73) |
| $Bi_2$-$O_1$ (×3) | 2.60 | 2.6472(70) |
| $Bi_2$-$O_3$ (×3) | 2.82 | 2.7968(99) |
| Ru-$O_1$ (×2) | 1.98 | 1.9201(10) |
| Ru-$O_3$ (×2) | 2.03 | 2.1907(92) |
| Ru-$O_3$ (×2) | 1.89 | 1.7477(11) |

### 3.2. Magnetic properties

Figure 2 shows the magnetic isotherms measured at different temperatures on a powder sample $Bi_{2.67}Pr_{0.33}Ru_3O_{11}$. At low temperature, the isotherms are curved but become linear as the temperature is increased above 20 K, as expected for an antiferromagnetic compound.

The thermal variation of the magnetic susceptibility in $Bi_{2.67}Pr_{0.33}Ru_3O_{11}$ (Figure 3a) was extracted from the measured magnetic isotherms by linear extrapolation to zero field. The results of similar measurements on the pyrochlore phase $Bi_2Ru_2O_7$ are included for comparison (Figure 3b), showing that the magnetic susceptibility in the latter is more than one order of magnitude smaller. This suggests that the major contribution in the $Bi_{2.67}Pr_{0.33}Ru_3O_{11}$ magnetization arises from $Pr^{3+}$ ions. The thermal variation of the magnetic susceptibility in $Bi_{2.67}Pr_{0.33}Ru_3O_{11}$ is also much lower than in unsubstituted $Bi_3Ru_3O_{11}$ [2] (about 0.00085 $\mu_B$/T f.u. at the lowest temperature).

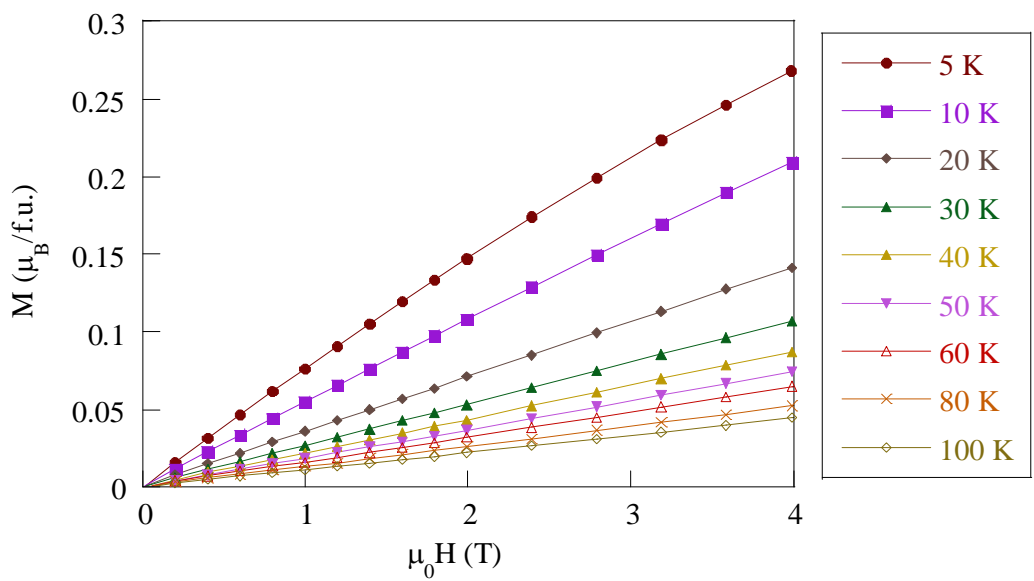

**Fig.2.** Magnetic isotherms measured at different temperature on a $Bi_{2.67}Pr_{0.33}Ru_3O_{11}$ powder sample.

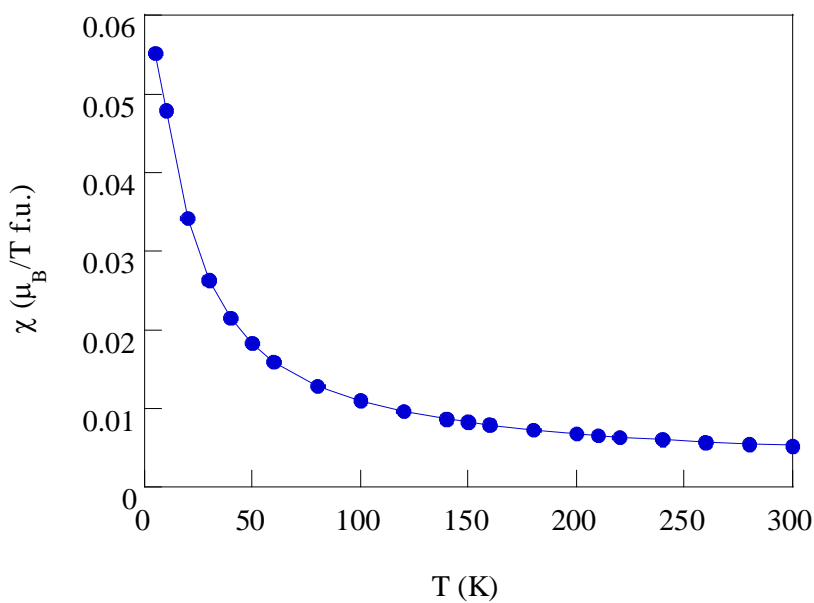

**Fig. 3.** Thermal variation of the magnetic susceptibility in $Bi_{2.67}Pr_{0.33}Ru_3O_{11.}$ (a) and in $Bi_2Ru_2O_{7.}$ (b).

The thermal variation of the reciprocal magnetic susceptibility is displayed in figure 4. It is markedly non linear at low temperature but tends to linear as the temperature increases. This non linear behavior should arise from the higher than quadrupolar order terms in the expansion of the crystal electric field interaction on the $Pr^{3+}$ ions, whose local symmetry is axial, more precisely $\bar{3}$ (three-fold roto inversion). The reciprocal magnetic susceptibility does not extrapolate to zero at zero temperature. However, it cannot be concluded that this would be due to some exchange interactions between the ions because a crystal electric field could induce a similar behaviour in a powder sample.

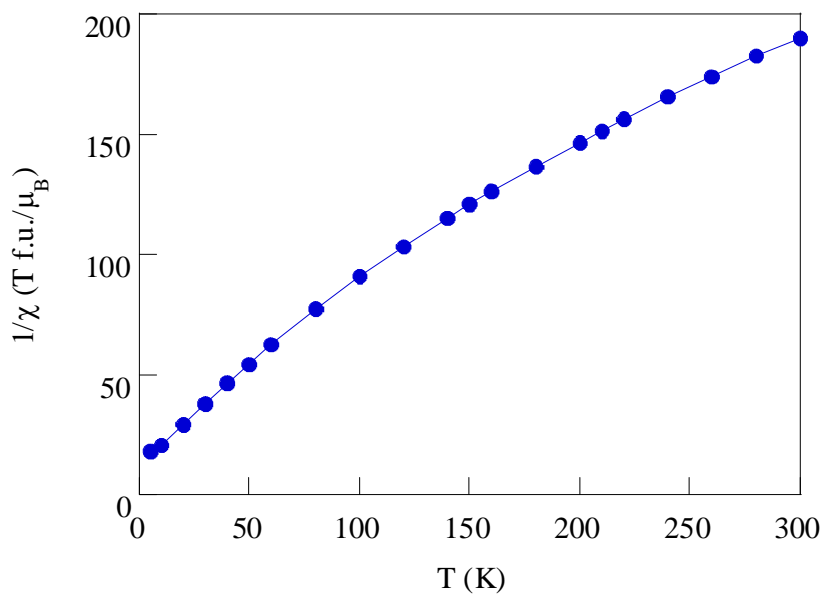

**Fig4.** Thermal variation of the reciprocal magnetic susceptibility on a $Bi_{2.67}Pr_{0.33}Ru_3O_{11}$ powder sample.

## Conclusions

A praseodymium-substituted phase in the Bi-Ru-O system is reported for the first time. The rare earth occupies the Bi1 site (4b) in the Pn3 structure and induces a cell volume decrease and shrinking in (Bi, Pr)-O average distance. Magnetic measurements show the dominant contribution of the $Pr^{3+}$ ions with crystal electric field interactions.